\documentstyle[emulateapj5,epsf]{aastex}

\def \farcs{\hbox{$.\!\!^{\prime\prime}$}}

\lefthead{Hoekstra et al.}
\righthead{Measurement of the bias parameter from weak lensing}
\begin{document}

\slugcomment{Draft for ApJ Letters} 

\title{Measurement of the bias parameter from weak lensing}

\author{Henk~Hoekstra$^{1,2,3}$, Howard K.C.~Yee$^{2,3}$, 
and Michael D.~Gladders$^{2,3}$}

\begin{abstract}
We have measured the correlation between the lensing signal induced by
(dark) matter and number counts of galaxies on scales ranging from
$0.15-3.0~h_{50}^{-1}$ Mpc (which correspond to aperture radii of 
$1-15$ arcminutes). This provides a direct probe of the scale
dependence of the ratio of the classical bias parameter $b$ and the
galaxy-mass correlation coefficient $r$. The results presented here
are based on 16 deg$^2$ of $R_C$ band data taken with the CFHT as part
of the Red-Sequence Cluster Survey. We used a sample of lens
galaxies with $19.5<R_C<21$, and a sample of source galaxies with
$21.5<R_C<24$. The results are consistent with a scale
independent value of $b/r$, which provides valuable constraints on
models of galaxy formation on scales that can only be probed through
weak lensing. For the currently favored cosmology $(\Omega_m=0.3,~
\Omega_\Lambda=0.7)$ we find $b/r=1.05^{+0.12}_{-0.10}$, similar to
what is found on larger scales ($\sim 10 h_{50}^{-1}$ Mpc) from local
dynamical estimates. These results support the hypothesis that light
traces mass on scales ranging from $0.15$ out to $\sim 10 h_{50}^{-1}$
Mpc. The accuracy of the measurement will improve significantly in the
coming years, enabling us to measure both $b$ and $r$ separately as a
function of scale.

\end{abstract}


\section{Introduction}

The weak distortions of the images of distant galaxies by intervening
matter (cosmic shear) provide an important tool to study the
projected mass distribution in the universe. The variance of the
lensing signal averaged in circular apertures can be used to constrain
cosmological parameters (e.g., Blandford et al. 1991; Kaiser 1992;
Schneider et al. 1998). Recently, various detections of this cosmic
shear signal have been reported (e.g., Bacon et al. 2000; Kaiser et
al. 2000; van Waerbeke et al. 2000; van Waerbeke et al. 2001; Wittman
et al. 2000), demonstrating the feasibility and prospects of this
technique.  

Other studies have concentrated on the dark matter halo properties of
galaxies (e.g., Brainerd et al. 1996; Hudson et al. 1998; Fischer et
al. 2000; Wilson et al. 2000; Hoekstra 2000).  Of these studies,
Fischer et al. (2000) were the first to measure the lensing signal
with high accuracy. They used the averaged tangential distortion
around their sample of lens galaxies to study the galaxy-mass
correlation function, which allowed them to investigate the ratio of
the bias parameter $b$ (e.g., Kaiser 1987) and the galaxy-mass
correlation coefficient $r$ (e.g., Pen 1998; Dekel \& Lahav 1999).

Other observational constraints of biasing come from dynamical
measurements (see Strauss \& Willick 1995) which provide a measurement
of the parameter $\beta=\Omega_m^{0.6}/b$ on relatively large scales
($\ge$ a few Mpc). Current measurements yield $\beta\sim 0.4$ (see the
compilation by Berlind et al. 2001; Peacock et al 2001) which suggest
for $\Omega_m\sim 0.3$ that $b/r\sim 1$.

\vbox{
\vspace{0.5cm}
\footnotesize
\noindent 
$^1$~CITA, University of Toronto,
Toronto, Ontario M5S 3H8, Canada\\
$^2$~Department of Astronomy, University of Toronto,
Toronto, Ontario M5S 3H8, Canada; hoekstra,gladders,hyee@astro.utoronto.ca\\
$^{3}$~Visiting Astronomer, Canada-France-Hawaii Telescope, which
is operated by the National Research Council of Canada, Le Centre 
National de Recherche Scientifique, and the University of Hawaii
}

The combination of high resolution cosmological $N$-body simulations
and semi-analytic modelling of galaxy formation has been used to
predict galaxy biasing as a function of scale (e.g., Benson et
al. 2000; Kauffmann et al. 1999). These studies find that on scales
$\le 4 h_{50}^{-1}$ Mpc the galaxies are less clustered than the dark
matter, although the results depend on the specific clustering
properties for different types of galaxies (e.g., Kauffmann et
al. 1999). A measurement of the galaxy biasing as a funcion of scale
is important as it can be used to rule out models of galaxy formation,
thus improving our understanding of this complex process.

Schneider (1998) proposed a method that provides the unique
opportunity to measure $b$ and $r$ as a function of scale. This has
been studied in more detail by Van Waerbeke (1998) who concluded that
the results depend only slightly on the assumed power spectrum of density
fluctuations.

In this letter we demonstrate the usefulness of weak gravitational
lensing for studies of galaxy biasing. We use $R_C$-band imaging data
from the Red-Sequence Cluster Survey (RCS) (e.g., Gladders \& Yee
2000), which is a 100 deg$^2$ galaxy cluster survey designed to
provide a large sample of optically selected clusters of galaxies with
redshifts $0.1<z<1.4$.

\section{Observations and analysis}

We use 16 deg$^2$ of $R_C$-band imaging data from the RCS taken with the
CFHT 12k mosaic CCD camera. The integration times are 900s per
pointing with seeing ranging from $0\farcs{6}$ to $0\farcs{9}$.  The
pipeline data reduction is described in detail in Gladders \& Yee
(2001). The CFHT part of the RCS consists of 10 widely separated
patches of $\sim 2.1\times 2.3$ degrees. Here we use 52 pointings of the
CFH12k camera, typically 5 pointings from each patch.  The data
analysed so far do not cover complete patches, and therefore we limit
the analysis to the individual pointings.

For the weak lensing analysis we use the scheme outlined in Hoekstra
et al. (1998; 2000), which is based on the method described in Kaiser,
Squires, \& Broadhurst (1995) and Luppino \& Kaiser (1997). A detailed
discussion of the object analysis and its accuracy is described in
Hoekstra et al. (2001), where we present our measurement of cosmic
shear. The results indicate that the object analysis, and the
necessary corrections for observational distortions work well, which
allows us to obtain accurate measurements of the weak lensing signal.

For the analysis presented here, we select a sample of lenses and
sources on the basis of their apparent $R_C$ magnitude. We use galaxies
with $19.5<R_C<21$ as lenses, and galaxies with $21.5<R_C<24$ as sources
which are used to measure the lensing signal. This selection yields a
sample of 36226 lenses and $\sim 6\times 10^5$ sources.

\section{Aperture mass and number counts}

To study the galaxy biasing as a function of scale, we essentially
measure the ratio of the cross-correlation between mass and galaxies,
and the galaxy auto-correlation function. The method is described in
detail in Schneider (1998) and Van Waerbeke (1998), and we discuss it
only briefly here. The aperture mass is defined as (Schneider et
al. 1998)

\begin{equation}
M_{\rm ap}(\theta)=\int d^2\phi~U(\phi)\kappa(\phi).
\end{equation}

\noindent If $U(\phi)$ is a compensated filter, i.e., $\int
d\phi~\phi~U(\phi)=0$, with $U(\phi)=0$ for $\phi>\theta$, the
aperture mass statistic $M_{\rm ap}$ can be directly expressed in terms
of the (in the weak lensing regime) observable tangential shear 
$\gamma_{\rm T}$

\begin{equation}
M_{\rm ap}(\theta)=\int d^2\phi~Q(|\phi|)\gamma_{\rm T}(\phi),
\end{equation}

\noindent where

\begin{equation}
Q(\theta)=\frac{2}{\theta^2}\int_{0}^{\theta} d\phi~\phi~U(\phi)~-~U(\theta).
\end{equation}

\noindent We will take (Schneider 1998; Van Waerbeke et al. 2001)

\begin{equation}
U(\phi)=\frac{9}{\pi \theta^2_{\rm ap}}
\left[1-\left(\frac{\phi}{\theta_{\rm ap}}\right)^2\right]
\left[\frac{1}{3}-\left(\frac{\phi}{\theta_{\rm ap}}\right)^2\right],
\end{equation}

\noindent with the corresponding $Q(\phi)$

\begin{equation}
Q(\phi)=\frac{6}{\pi\theta^2_{\rm ap}}
\left(\frac{\phi}{\theta_{\rm ap}}\right)^2
\left[1-\left(\frac{\phi}{\theta_{\rm ap}}\right)^2\right].
\end{equation}

The real bias relation can be complicated as it depends on the process
of galaxy formation (e.g., Dekel \& Lahav 1999), but in the standard,
deterministic, linear bias theory, the galaxy density contrast
$\delta_{\rm g}$ is related to the mass density contrast $\delta$ as
(e.g., Kaiser 1987) $\delta_{\rm g}=b \delta$, and the number density
contrast of galaxies is given by

\begin{equation}
\Delta n_{\rm g}(\theta)=\frac{N(\theta)-\bar N}{\bar N}=
b \int dw p_f(w) \delta(f_K(w)\theta;w),
\end {equation}

\noindent where $\bar N$ is the average number density of lens
galaxies, $w$ is the comoving distance, $f_K(w)$ is the comoving
angular diameter distance, and where $p_f(w)dw$ corresponds to the
redshift distribution of lens galaxies.

We define the filtered number counts ${\cal N}(\theta_{\rm ap})$ as
(Schneider 1998)

\begin{equation}
{\cal N}(\theta_c)=\int d^2\phi U(\phi)\Delta n_{\rm g}(\phi).
\end{equation}

For our choice of the filter function $U(\phi)$ we define the
`filtered' power spectrum $P_{\rm filter}(w;\theta_{\rm ap})$ as
(e.g., Schneider et al. 1998)

\begin{equation}
P_{\rm filter}(w;\theta_{\rm ap})\hspace{-0.3mm}=\hspace{-1.2mm}\int\hspace{-1.0mm}
ds~sP_{\rm 3d}\hspace{-0.3mm}\left(\frac{s}{f_K(w)};w\right)\hspace{-1.5mm}
\left[\frac{12 J_4(s \theta_{\rm ap})}
{\pi (s \theta_{\rm ap})^2}\right]^2,
\end{equation}

\noindent where $P_{\rm 3d}$ is the time-evolving 3-D power spectrum,
and $J_4(x)$ is the fourth order Bessel function of the first kind.
As has been shown by Jain \& Seljak (1997), it is important to use the
non-linear power spectrum in the calculations, for which the results
from Peacock \& Dodds (1996) are used.

We can write the auto-correlation of $\cal N$ as (Schneider 1998; Van
Waerbeke 1998)

\begin{equation}
\langle{\cal N}^2(\theta_c)\rangle=2\pi b^2
\int dw \frac{p_f^2(w)}{f_K^2(w)} P_{\rm filter}(w;\theta_{\rm ap}),
\end{equation}

\noindent and the cross-correlation $\langle M_{\rm ap}(\theta_{\rm ap})
{\cal N}(\theta_{\rm ap})\rangle$ as

\begin{equation}
\begin{array}{rl}
\langle M_{\rm ap}(\theta_c){\cal N}(\theta_c)\rangle  &  
= 3\pi \left(\frac{H_0}{c}\right)^2 \Omega_m b r\times\\
  & \int{dw \frac{p_f(w) g(w)}{a(w) f_K(w)} P_{\rm filter}(w;\theta_c)},\\
\end{array}
\end{equation}

\noindent where $\Omega_m$ is the density parameter, and $a$ is the
cosmic expansion factor. 

Schneider (1998) and Van Waerbeke (1998) used the simple,
deterministic, linear biasing theory, which assumes a perfect correlation
between the galaxy and mass density fields. However, as pointed out by Pen
(1998), and Dekel \& Lahav (1999), the biasing relation need not be
deterministic, but might be stochastic. In this case the galaxy-mass 
cross-correlation coefficient $r$ is less than 1. Hence, we allow
for stochastic biasing, and include the parameter $r$.

The function $g(w)$ depends on the redshift distribution of
the (background) sources $p_b(w)dw$ as

\begin{equation}
g(w)=\int_{w}^{w_H} dw' p_b(w') \frac{f_K(w'-w)}{f_K(w')}.
\end{equation}

The predicted correlations $\langle{\cal N}^2\rangle$ and 
$\langle M_{\rm ap}{\cal N}\rangle$ depend on the choice 
of the power spectrum. Van Waerbeke (1998) showed that
the ratio

\begin{equation}
{\cal R}(\theta_{\rm ap})=\frac{\langle M_{\rm ap}(\theta_{\rm ap}) 
{\cal N}(\theta_{\rm ap})\rangle} {\langle{\cal N}^2(\theta_{\rm ap})\rangle},
\end{equation}

\noindent depends little on the choice of power spectrum, provided
that $b$ and $r$ are independent of scale. In this case, the ratio
only depends on $\Omega_m$ and $\Omega_\Lambda$, and is constant with
scale\footnote{We found that the ratios presented in Figure~5 from 
Van Waerbeke (1998) are in fact proportional to $\sqrt{{\cal R}b/r}$, 
which obviously does not alter the conclusion that ${\cal R}$ is constant with scale}.
Thus we can write

\begin{equation}
{\cal R}=\Omega_m \frac{r}{b} \times f(\Omega_m,\Omega_\Lambda),
\end{equation}

\noindent where $f(\Omega_m,\Omega_\Lambda)$ is a constant for a given
cosmology. Thus the measurement of ${\cal R}$ as a function of scale
provides a unique way to examine whether $b/r$ depends on scale or
not. In fact, weak gravitational lensing allows one to estimate $b$
and $r$ separately, when the ratio ${\cal B}={\langle M_{\rm
ap}^2\rangle}/ {\langle{\cal N}^2\rangle}\propto(\Omega_m/b)^2$ is
also measured. Although this ratio is not constant with scale, ${\cal
B}$ does not depend much on the assumed power spectrum (in particular
on scales less than 10 arcmin.). Unfortunately our data are not
sufficient to obtain a good measurement of $\langle M_{\rm
ap}^2\rangle$, but with more data we will be able to measure both $r$
and $b$ as a function of scale.

\vbox{
\begin{center}
\leavevmode 
\hbox{%
\epsfxsize=\hsize 
\epsffile[5 145 580 710]{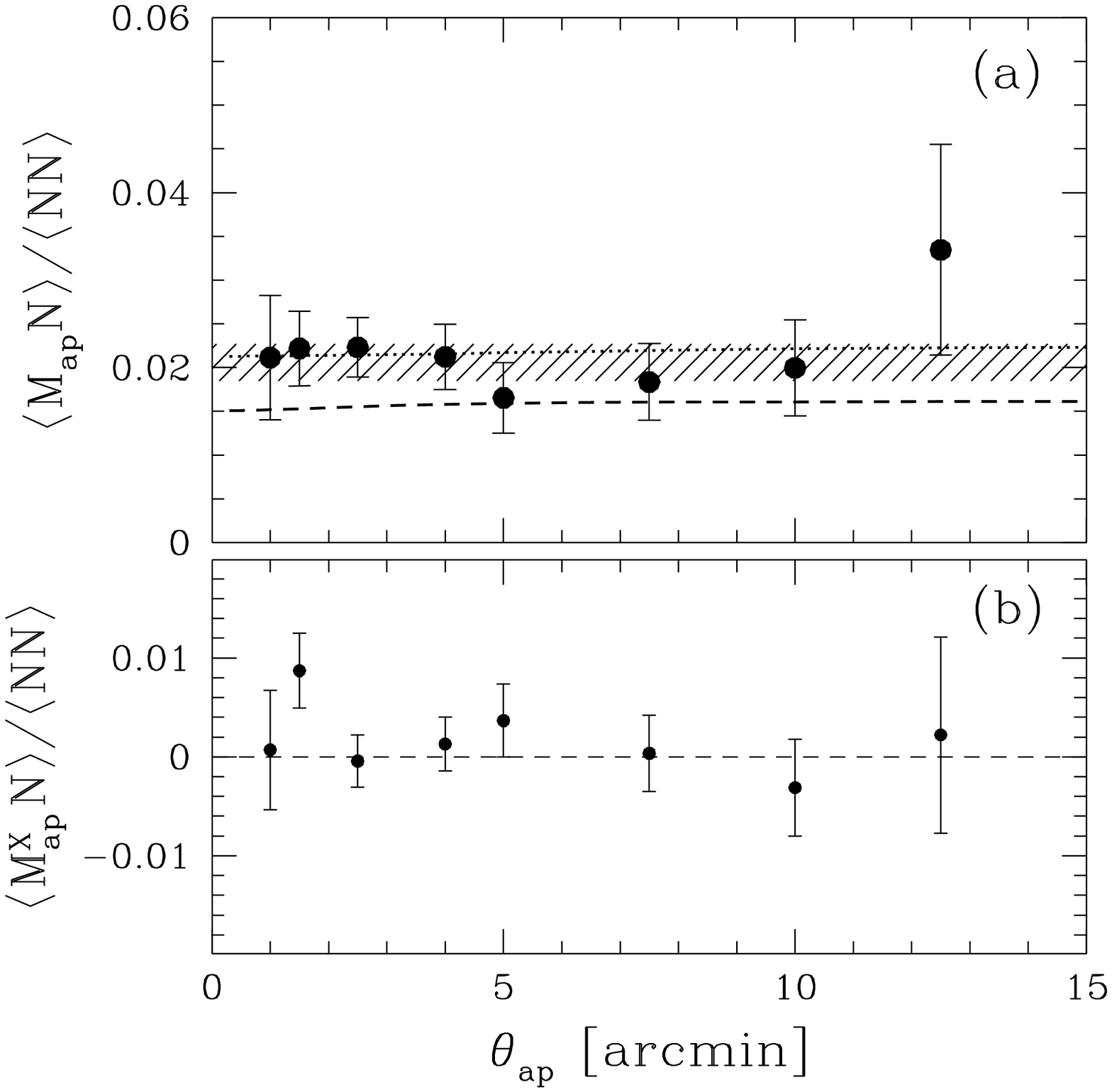}} 
\figcaption{\footnotesize (a) The observed ratio of
${\cal R}$ as a function of aperture radius $\theta_{\rm ap}$. Note
that the points are somewhat correlated. The errorbars are computed
using the scatter in the measurements of the individual fields. 
The dashed line indicates the model predictions for an OCDM model, and the
dotted line corresponds to an LCDM model. We find an average value of
${\cal R}=0.021\pm0.002$ (indicated by the hatched region), where we 
have used the full covariance matrix in order to account for the 
correlation between the points. (b) The measurement when the phase of 
the shear is increased by $\pi/2$, which should vanish if the signal in 
(a) is caused by lensing. The results are indeed consistent with no signal. 
The signal also vanishes when we correlate $\cal N(\theta_{\rm ap})$ of a 
given field with $M_{\rm ap}(\theta_{\rm ap})$ measured from the other 
pointings.
\label{ratio}}
\end{center}}

\section{Redshift distributions}

In order to interpret the observed value of ${\cal R}$, we have to
evaluate equations~9 and~10, which requires knowledge of the
redshift distributions of the lens galaxies and the source
galaxies.

For the sample of lens galaxies we use the redshift distribution
found from the CNOC2 Field Galaxy Redshift Survey (e.g., Lin et
al. 1999; Yee et al. 2000; Carlberg et al. 2000).  The CNOC2 survey
provides a well determined (spectroscopically) redshift distribution
for field galaxies down to $R_C=21.5$, which is ideal, given our
limits of $19.5<R_C<21$. The adopted redshift distribution gives a
median redshift $z=0.35$ for the lens galaxies.

For the source galaxies the situation is more complicated.
These galaxies are generally too faint for spectroscopic surveys,
although recently Cohen et al. (2000) measured spectroscopic redshifts
around the Hubble Deep Field North down to $R_C\sim 24$. Cohen et al. (2000)
find that the spectroscopic redshifts agree well with the photometric
redshifts derived from multi color photometry. Because of likely
field-to-field variations in the redshift distribution, we prefer
to use the photometric redshift distributions derived from 
both Hubble Deep Fields (Fern{\'a}ndez-Soto et al. 1999; Chen et al. 1998). 
Photometric redshift distributions generally work well, as
has been demonstrated by Hoekstra et al. (2000). This redshift
distribution yields a median redshift of $z=0.53$ for the source galaxies.

We computed the value of ${\cal R}$ for a range of cosmological parameters,
and find that, for the adopted redshift distributions, ${\cal R}$ can
be approximated with fractional accuracy of 2\% using

\begin{equation}
{\cal R}= \frac{r\Omega_m}{100 b}\left[(5.8-1.6\Omega_m^{0.63})+ 
 (4.6-2.6\Omega_m^{0.63})\Omega_\Lambda^{1.23}\right].
\end{equation}

\section{Measurement of the bias parameter}

To measure $\langle M_{\rm ap}{\cal N}\rangle$ and $\langle {\cal N}^2\rangle$
from the data we use the estimators for $M_{\rm ap}$ and ${\cal N}$
introduced by Schneider (1998)

\begin{equation} 
\tilde M_{\rm ap}= \pi\theta_{\rm ap}^2
\frac{\sum_{i=1}^{N_b} Q(\theta_i) w_i \gamma_{{\rm T},i}}
{\sum_{i=1}^{N_b} w_i},~{\rm and}~
\tilde {\cal N}=\frac{1}{\bar N} \sum_{i=1}^{N_f} U(\theta_i),
\end{equation}

\noindent where $N_f$, and $N_b$ are respectively the number of
lens and source galaxies found in the aperture of radius
$\theta_{\rm ap}$. The weights $w_i$ correspond to the inverse square 
of the uncertainty in the shape measurement (see Hoekstra et al. 2000
for a detailed discussion).

The observed value of $\cal R$ as a function of aperture size is
presented in Figure~\ref{ratio}a. We note that the points are somewhat
correlated. A significant signal is detected at all scales. The
results are consistent with a value of ${\cal R}$ that is constant
with scale, which implies that $b/r$ is constant as well. This is an
important result, as the smallest scales we are probing are comparable
to the sizes of galaxy halos. We obtain an average value of ${\cal
R}=0.021\pm0.002$, where we have used the covariance matrix to account
for the correlation between the points at different scales.

To examine possible systematic effects, we also computed $\langle
M_{\rm ap}{\cal N}\rangle$ when the galaxies are rotated by 45
degrees. This signal should vanish in the case of lensing. The results
presented in Figure~\ref{ratio}b are consistent with no signal,
indicating that the corrections for the systematic distortions have
worked well (more details will be provided in Hoekstra et
al. 2001). As another check, we correlated ${\cal N}(\theta_{\rm ap})$
for each field with $M_{\rm ap}(\theta_{\rm ap})$ of the other
pointings, and find that the signal also vanishes in this case.

\vbox{
\begin{center}
\leavevmode 
\hbox{%
\epsfxsize=\hsize 
\epsffile[30 287 590 700]{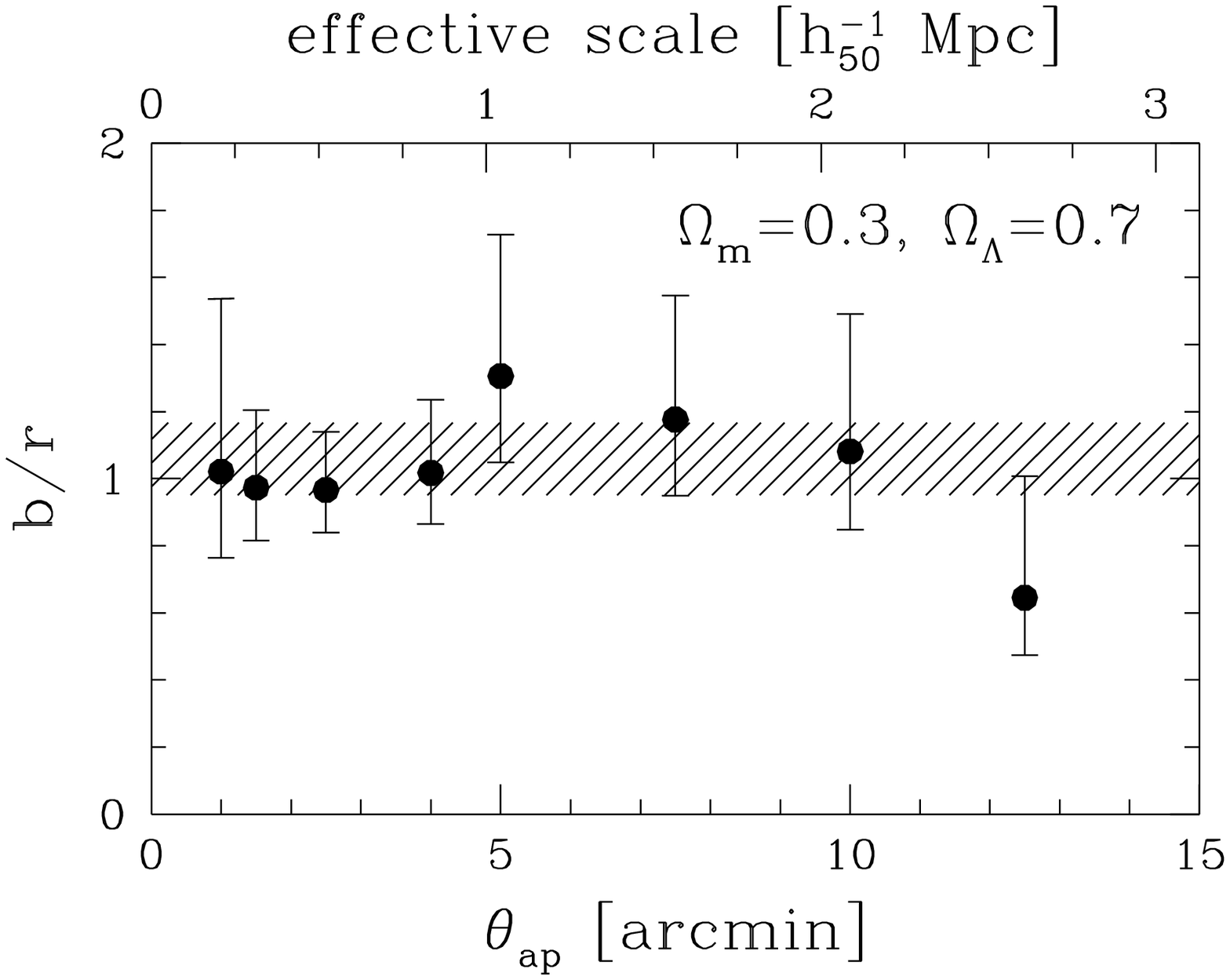}} 
\figcaption{\footnotesize The value of $b/r$ as a function of angular
scale, under the assumption $\Omega_m=0.3$ and
$\Omega_\Lambda=0.7$. Note that the points are slightly correlated.
The errorbars (which indicate the 68\% confidence limits) are computed
using the scatter in the measurements of the individual fields.  The
upper axis indicates the effective physical scale probed by the
compensated filter $U(\phi)$ at the median redshift of the lenses
$(z=0.35)$. The results are consistent with a
value of $b/r$ that is independent of scale. For this cosmology we
find $b/r=1.05^{+0.12}_{-0.10}$ (indicated by the hatched region),
whereas for an open model $(\Omega_m=0.3,~\Omega_\Lambda=0.0)$ we
obtain $b/r=0.73^{+0.08}_{-0.07}$. The error on the average has been
computed using the full covariance matrix, in order to account for the
correlation between the points at various scales.
\label{bias}}
\end{center}}

In Figure~\ref{bias} we present the resulting value of $b/r$ as a
function of aperture radius for the currently favored cosmology
$(\Omega_m=0.3,~\Omega_\Lambda=0.7)$. In this case we find
$b/r=1.05^{+0.12}_{-0.10}$. For an open model
$(\Omega_m=0.3,~\Omega_\Lambda=0.0)$ we obtain
$b/r=0.73^{+0.08}_{-0.07}$. For comparison, we have also indicated the
effective physical scale ($\sim$ FWHM of the filter function) probed
by the compensated filter $U(\phi)$ at the median redshift of the
lenses $(z=0.35)$.

A direct comparison with dynamical studies is difficult because
different galaxy types cluster differently and because of the
different scales probed in our study. However, our results are in fair
agreement with the results from dynamical studies, in the sense that
we find $b/r \sim 1$ (e.g., Berlind et al. 2001; Peacock et
al. 2001). Therefore from scales ranging from $0.15 h_{50}^{-1}$ Mpc
out to $\sim 10 h_{50}^{-1}$ Mpc, i.e. from the scales of galaxy halos
out to the linear regime, the measurements are consistent with a value
$b/r \sim 1$, suggesting that the light distribution traces the dark
matter distribution quite well. 

\section{Prospects}

For the first time we have measured the parameter $b/r$ as a function
of scale using weak lensing using 16 deg$^2$ of data from the
Red-Sequence Cluster Survey. With the analysis of the full survey, the
errorbars are expected to decrease by a factor $\sim 2$, thus
improving the constraints on possible variation of $b/r$ with scale.
Also we will be able to probe larger scales, as we have limited the
analysis to the individual pointings, rather than the full patches
which are $\sim 2.1\times 2.3$ degrees. Other cosmic shear surveys will
place additional constraints, eventually allowing us to measure $r$
and $b$ separately as a function of scale.

The lens galaxies were selected on the basis of their apparent
magnitude, but with planned multi-color photometry, it is also
possible to measure the biasing properties as a function of galaxy
type or luminosity (using photometric redshifts). Eventually, using
bigger surveys, it might even be possible to study the evolution of
galaxy biasing as a function of redshift.

\end{document}